\documentclass{amsart}
\usepackage{graphicx}
\usepackage{amsmath}
\theoremstyle{definition}

\theoremstyle{remark}

\numberwithin{equation}{section}
\newcommand{\norm}[1]{\left\Vert#1\right\Vert}

\newtheorem{example}{Example}

\def\ket#1{| #1 \rangle}
\def\ss{{\rm ss}}
\def\DD{\mathfrak{D}}

\def\vec#1{{\bf #1}}
\begin{document}
\title[Quantum Control of Two-Qubit Entanglement Dissipation]{Quantum Control of Two-Qubit Entanglement Dissipation}%
\author{Sophie G. Schirmer}%
\address
{School of Physical Sciences,
Swansea University,
Singleton Park,
Swansea SA2 8PP, Wales UK
}%
\email{sgs29@swansea.ac.uk }%
\author{Allan I. Solomon}
\address{Department of Physics and Astronomy, The Open University, MK7 6AA, UK \\and \\LPTMC, Universit\'e  de Paris VI, France}
\email{a.i.solomon@open.ac.uk}

\begin{abstract}
We investigate quantum control of the dissipation of entanglement under
environmental decoherence. We show by means of a simple two-qubit model that standard control methods - coherent or open-loop control - will not in general  prevent  entanglement loss.  However, we propose a control method utilising a Wiseman-Milburn feedback/measurement control scheme which will effectively negate environmental entanglement dissipation.
\end{abstract}
\maketitle
\section{Introduction}

Entanglement has recently emerged as a significant resource in quantum
information and other applications.  However, an important problem is to
ensure the robustness of this resource; that is, the ability to maintain
it against decay.  Unlike other quantum properties, entanglement is not
invariant under general unitary transformations, and this implies the
possibility of decay under unitary forces.  These are however,
reversible and thus conversely enable entanglement production, given
appropriate control methods.

Additionally, environmental dissipation is an ever-present source of the
loss of entanglement; and it is therefore important to devise quantum
control procedures to protect against this loss, where possible.

In this note we analyze the possibility of preservation of entanglement
against decay for a simple two-qubit Bell state by means of quantum
control.  We conclude that this is not possible for a simple hamiltonian
quantum control process (open-loop control) but show that an appropriate
feedback/measurement control procedure can prove effective protection
against environmental loss in certain cases.

\section{Unitary Dissipation}

Quantum dissipation is usually associated with non-unitary,
non-reversible processes. However {\em entanglement} is subject to
unitary dissipation, since unitary evolution associated with a
(hermitian) hamiltonian does not necessarily preserve entanglement.
Conversely, entanglement may be {\em produced} by the evolution induced
by a quantum control hamiltonian.  We choose a simple example to
illustrate the bipartite case.

\subsection{Entanglement production}
\label{entanglement}

Consider the unitary evolution $U(t)$ induced by the hamiltonian $H$
given by
\begin{equation}
 \label{ham}
   H= \begin{pmatrix}
       {\it x_1}&0&0&0\\\noalign{\medskip}
       0&{\it x_2}&y&0\\\noalign{\medskip}
       0&y&{\it x_3}&0\\\noalign{\medskip}
       0&0&0&{\it x_4}
       \end{pmatrix},
\end{equation}
which corresponds to a system with Heisenberg coupling and local
control terms
\begin{equation}
  H = a Z \otimes I + b I\otimes Z +
c (X \otimes X + Y \otimes Y + Z \otimes Z)
\end{equation}
with $x_1=a+b+c$, $x_2=a-b-c$, $x_3=-a+b-c$, $x_4=-a-b+c$, $y=2c$.  For
simplicity we  assume $a=b$ so that $x_2=x_3$.

We shall use the Concurrence ${\mathcal C}$~\cite{woot} as a measure of
entanglement for a bipartite two-qubit system.  The {\em concurrence}
${\mathcal C}$ of a two-qubit state $\rho$ is given by
\begin{equation}
 \label{eq:c1}
 {\mathcal C} = \max\left\{\lambda _1-\lambda _2-\lambda _3-\lambda_4,0\right\},
\end{equation}
where the quantities $\lambda _i$ are the square roots of the
eigenvalues of the $4 \times 4$ matrix
\begin{equation}
  \label{eq:c2}
  \rho (Y\otimes Y)\rho^{*}(Y\otimes Y)
\end{equation}
in descending order, where $Y=\begin{bmatrix} 0&-i\\i&0 \end{bmatrix}$.

Acting with the unitary evolution matrix $U(t)=\exp(itH)$ induced by
Eq.(\ref{ham}) on the base vector $\vec{v_0}\equiv [0,1,0,0]^T$ gives
\footnote{Note that for such calculations it is important to choose a
fixed basis - here we choose the standard basis.}
\begin{align*}
 v(t)&= \exp{(itH)}v_0 \\
     &=e^{it x_2}\, [0,\cos(ty),i\sin(ty),0]^T,
\end{align*}
from which we can easily obtain the concurrence of $v(t)$ to be
$|\sin(2ty)|$.

\subsection{Unitary dissipation of Entanglement}

The off-diagonal coupling term $y$ changes the entanglement.  This
can be used to create entanglement but it also destroys entanglement.

Referring to Figure 1, we see immediately that if we start in the state
$\vec{v_0}$ then at $t=\pi/4$ (in units of $1/y$) we have the maximally
entangled (Bell) state $\tfrac{1}{\sqrt{2}}[0,1,i,0]^T$ (up to a global
phase factor) but the unitary action $U(t)$ destroys the entanglement,
completely at $t=\pi/2$.  

\begin{figure}
\vspace{1cm}
\begin{center}\resizebox{8 cm}{!}
{\includegraphics{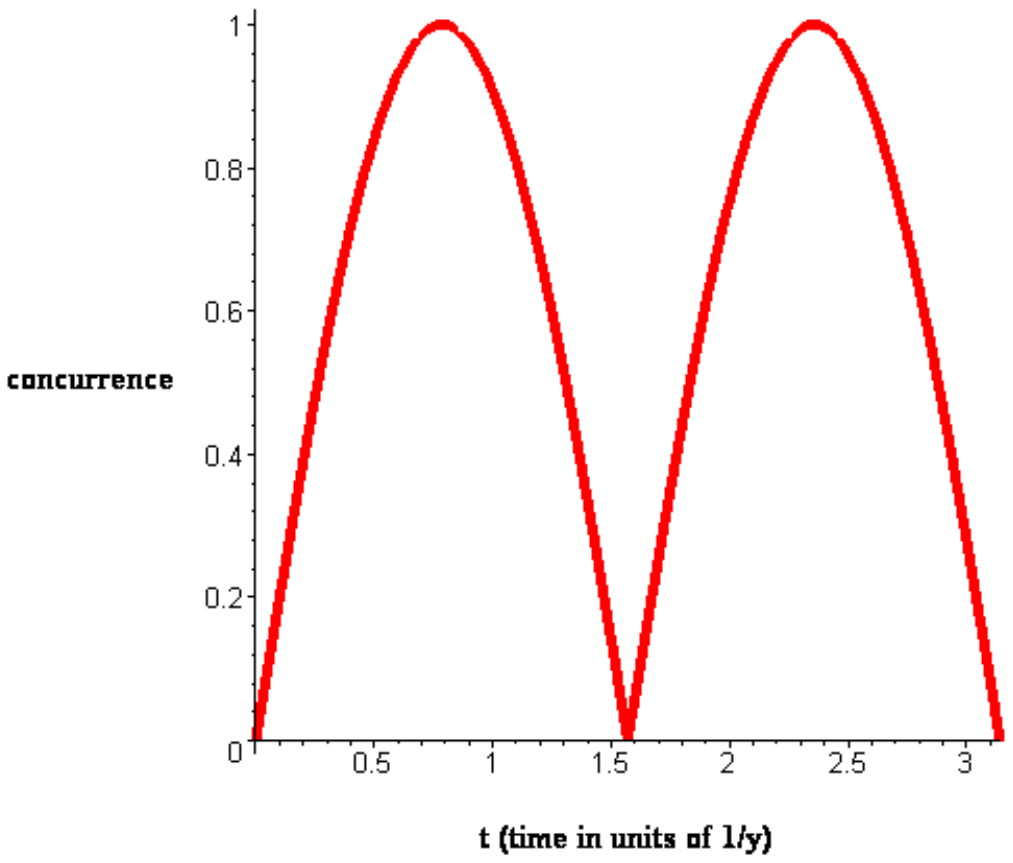}}
\caption{Concurrence vs Time $t$ (units of $1/y$)}
\label{plot1}
\end{center}
\end{figure}

\section{Environmental Dissipation}
\label{envir}

The standard form of a general Markovian {\em dissipative} process in
Quantum Mechanics is governed by the Liouville equation obtained by
adding a dissipation (super-)operator ${L}_D[{\rho}(t)]$ to the usual
hamiltonian term:
\begin{equation}
   \label{eq:dLE}
   \dot{\rho}(t) = -i[{H},{\rho}(t)] + L_D[{\rho}(t)]
\end{equation}
where the density matrix $\rho$ for an $N$-level system is an $N \times
N$ (semi-)positive matrix.

\subsection{Lindblad Equation}

In order that the quantum Liouville equation Eq.(\ref{eq:dLE}) should
define a {\em physical} dissipation process, certain constraints must be
imposed, most notably complete positivity.  The appropriate constraints
emerge from physical stochastic dissipation equations such as those
given by Lindblad and others in differential form \cite{Lindblad}, as
well as in global form \cite{Kraus}.  The result is that completely
positive evolution of the system is guaranteed provided the dissipation
super-operator $L_D$ takes following form
\begin{equation}
\label{lin}
  L_D[{\rho}(t)] =
  \frac{1}{2} \sum_{s=1}^{N^2} \left\{[{V}_s{\rho}(t),{V}_s^\dag] +
                               [{V}_s,{\rho}(t){V}_s^\dag] \right\}
\end{equation}
where the matrices $V_s$ are arbitrary.  Denoting the standard basis for
$N \times N$ matrices by
\begin{equation}
\label{sb}
 (E_{ij})_{mn}={\delta}_{im}{\delta}_{jn} \qquad (i,j,m,n=1 \ldots N)
\end{equation}
and relabelling, using the notation $[m;n] = (m-1)N+n$, we have
\begin{equation}
  \label{av}
  V_{[i;j]}=a_{[i;j]}E_{[i;j]}.
\end{equation}
The virtue of Eq.~(\ref{eq:dLE}) is that every Markovian dissipation
process has to satisfy it and so results derived from its use have
great generality.  A drawback, however, is that it is often not obvious
how to relate phenomenological observations of dissipative effects to
dissipation operators of the form (\ref{lin}).

\subsection{Phenomenological description}

Having chosen a certain computational or preferred basis, one can
phenomenologically distinguish two types of dissipation: phase
decoherence and population relaxation.  The former occurs when the
interaction with the environment destroys the phase correlations of
certain superposition states.  In the simplest case this leads to a
decay of the diagonal elements $\rho_{kn}(t)$ of the density operator at
a constant (dephasing) rate $\Gamma_{kn}$:
\begin{equation}
\label{eq:dephasing}
 \dot{\rho}_{kn}(t)
  = -i([{H},{\rho}(t)])_{kn}-\Gamma_{kn}\rho_{kn}(t).
\end{equation}
Population relaxation occurs, for instance, when a quantum particle in a
certain state spontaneously emits a photon and transitions to a less
energetic quantum state.  In the simplest case, when there are only jumps
between the basis states, $\ket{k}$ and $\ket{n}$ say, occurring at
fixed rates $\gamma_{kn}$, the resulting population changes can be
modelled as
\begin{equation}
 \label{eq:poptrans}
 \dot{\rho}_{nn}(t)
 = -i([{H},{\rho}(t)])_{nn}
 +\sum_{k\neq n} \left[\gamma_{nk}\rho_{kk}(t)-\gamma_{kn}\rho_{nn}(t)\right]
\end{equation}
where $\gamma_{kn}\rho_{nn}$ is the population loss for level $\ket{n}$
due to transitions $\ket{n}\to\ket{k}$, and $\gamma_{nk}\rho_{kk}$ is
the population gain caused by transitions $\ket{k}\to\ket{n}$.  The
population relaxation rate $\gamma_{kn}$ is determined by the lifetime
of the state $\ket{n}$ and, for multiple decay pathways, the relative
probability for the transition $\ket{n}\to\ket{k}$.

Phase decoherence and population relaxation lead to a dissipation
super-operator (represented by an $N^2 \times N^2$ matrix) whose
non-zero elements are
\begin{subequations}
\label{supop}
  \begin{align}
  ({L}_D)_{[k;n],[k;n]}  &= -\Gamma_{kn} & k \neq n \\
  ({L}_D)_{[n;n],[k;k]}  &= +\gamma_{nk} & k \neq n \\
  ({ L}_D)_{[n;n],[n;n]} &= -\sum_{n\neq k} \gamma_{kn}
 \end{align}
\end{subequations}
where $\Gamma_{kn}$ and $\gamma_{kn}$ are taken to be positive numbers,
with $\Gamma_{kn}$ symmetric in its indices, and again we employ the
convenient notation $[m;n] = (m-1)N+n$ introduce above.

The $N^2 \times N^2$ matrix super-operator $L_D$ may be thought of as
acting on the $N^2$-vector $\vec{r}$ obtained from $\rho$ by
\begin{equation}
 \label{vec}
  \vec{r}_{[m;n]} \equiv \rho_{mn}.
\end{equation}
The resulting vector equation is
\begin{equation}
 \label{veceq}
  \dot{\vec{r}} = L \vec{r} = (L_H+L_D) \vec{r}
\end{equation}
where $L_H$ is the anti-hermitian matrix corresponding to the
hamiltonian $H$.  We obtain $L_H$ explicitly by using the standard
algebraic trick applied in evaluating Liouville equations (see, for
example~\cite{havel}).  The correspondence between $\rho$ and $\vec{r}$
as given in Eq.~(\ref{vec}) tells us, after some manipulation of
indices, that
\begin{equation}
 \label{trick}
   \rho \to \vec{r}
   \Rightarrow A \rho B \to A \otimes \tilde{B} \vec{r}
\end{equation}
using the direct (Kronecker) product of matrices.

\subsection{Constraints on Dephasing Rates}
\label{phys}

The phenomenological description above does not impose any constraints
on the population relaxation and decoherence parameters present in the
dissipation matrix.  In practice, however, the {\em values} of the
dissipation parameters $\Gamma_{kn}$ and $\gamma_{kn}$ must satisfy
various constraints~\cite{constraints} to ensure that they describe {\em
physical} processes, which can be derived by use of the Lindblad equation.

For simplicity we restrict ourselves here to the case of pure dephasing.
Experimentally, this is often the dominant decoherence process as the
population relaxation (or $T_1$) times for most systems are much longer
than the dephasing (or $T_2$) times so that we may effectively neglect
the relaxation rates $\gamma$.  In the {\em pure decoherence
(dephasing)} case, comparison of Eq.(\ref{lin}) and Eq.(\ref{av}) with
Eq.(\ref{supop}) tells us that the $\gamma$ terms vanish if we choose
$a_{[i;j]}=0$ for $i\neq j$.  The decoherence parameters $\Gamma_{ij}$
are then given by
\begin{equation}
  \label{Gammas}
 \Gamma_{ij} = \frac{1}{2}(|a_{[i;i]}|^2 + |a_{[j;j]}|^2)\; \; \; \; (i,j=1 \ldots N\;\;\;i \neq j).
\end{equation}

This leads to a mathematically very simple situation, as the dissipation
matrix $L_{D0}$ is then diagonal.  For the $N=4$ system, this gives six
pure dephasing parameters ($\Gamma_{ij}=\Gamma_{ji}$, $\Gamma_{ii}=0$),
determined by four constants, so there are two relations between the
$\Gamma$'s.  Explicitly, we have for the two-qubit, 4-level case,
\begin{subequations}
\label{LD0}
\begin{align}
  L_{D0}& ={\rm diag} \{0,-\Gamma_{{12}},-\Gamma_{{13}},-\Gamma_{{14}},-\Gamma_{{21}},0,-\Gamma_{{23}},-\Gamma_{{24}}, \nonumber \\
        &-\Gamma_{{31}},-\Gamma_{{32}},0,-\Gamma_{{34}},-\Gamma_{{41}},-\Gamma_{{42}},-\Gamma_{{43}},0
\}
\end{align}
\end{subequations}
with $\Gamma_{ij}=\Gamma_{ji}$, and the constraints that must be imposed
to ensure that we have physical process are then
\begin{equation}
 \label{Gs}
  \Gamma_{12}+\Gamma_{34}=\Gamma_{14}+\Gamma_{23}=\Gamma_{13}+\Gamma_{24}.
\end{equation}

\subsection{Markovian dissipation of entanglement}

We now give an example of a standard dephasing process acting on a
maximally entangled state.

\begin{example}
\label{bcoh}
Consider the Bell state
\begin{equation}
 \label{bell}
  v_B= \tfrac{1}{\sqrt{2}}[0,1,1,0]^T.
\end{equation}
The corresponding Liouville vector is
\(
  \vec{r} = \frac{1}{2}
  [0,0,0,0,0,1,1,0,0,1,1,0,0,0,0,0]^T
\)
and the action of the dephasing operator $L_{D0}$ of Eq.(\ref{LD0})
is given by
 \begin{equation}
  \label{veceq0}
 \dot{\bf{r}} = L {\bf{r}} = (L_{D0}) {\bf{r}},
\end{equation}
as in Eq.~(\ref{veceq}).  This equation may be immediately integrated to give
\begin{equation}
   \vec{r}(t) = \frac{1}{2}
  [0,0,0,0,0,1,e^{-\Gamma_{23}t},0,0,e^{-\Gamma_{23}t},1,0,0,0,0,0]^T
\end{equation}
corresponding to the density matrix
\begin{equation}
 \rho(t)= \frac{1}{2} \begin{pmatrix}
              0&0&0&0\\
	      0&1&e^{-\Gamma_{23}t}&0\\
	      0&e^{-\Gamma_{23}t}&1&0\\
	      0&0&0&0
	     \end{pmatrix}.
\end{equation}
Note that this does not represent a pure state except at $t=0$.  The
concurrence as defined in Section \ref{entanglement} evaluates to
$\exp(-\Gamma_{23}t)$. (See Figure 2).
\end{example}

\begin{figure}
\label{plot2}
\vspace{1cm}
\begin{center}\resizebox{8 cm}{!}
{\includegraphics{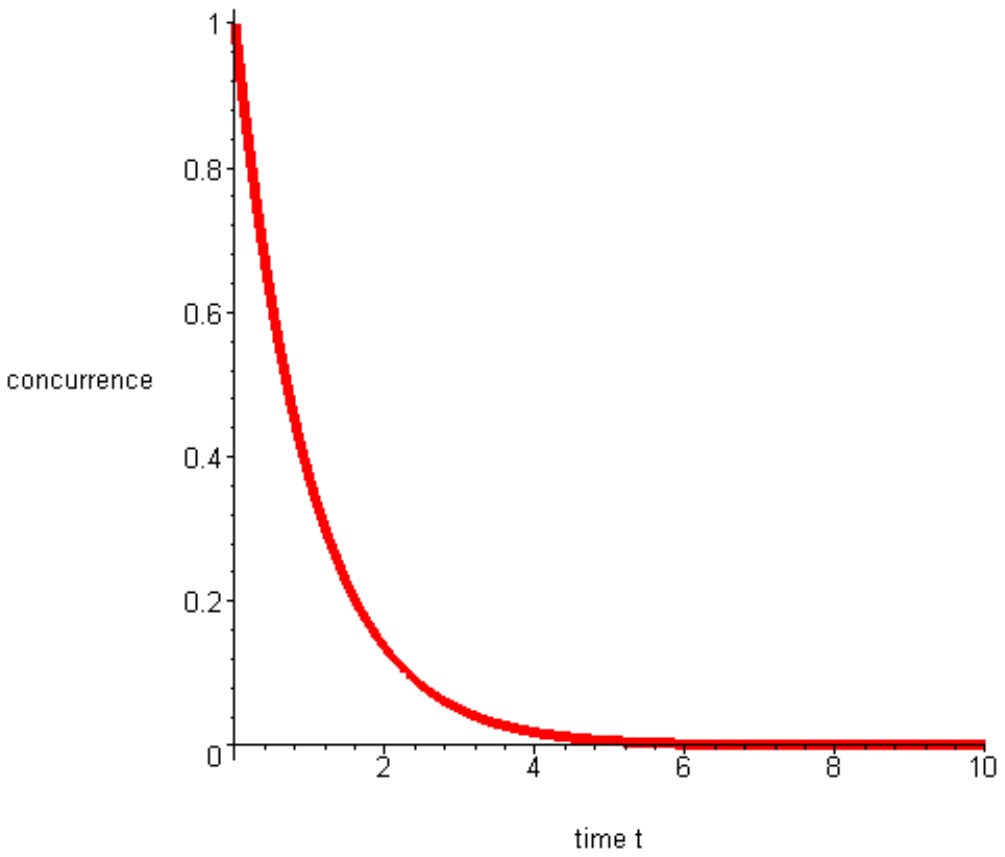}}
\caption{Concurrence vs Time $t$ (units of $1/\Gamma_{23}$)}
\end{center}
\label{conplot}
\end{figure}

The results of Example \ref{bcoh} are essentially unchanged in the
presence of an additional {\em free (i.e. diagonal)} hamiltonian as this commutes
with the dissipation super-operator $L_{D0}$\cite{sol}.

\section{Quantum control of entanglement dissipation}

In this section we analyze two quantum control schemes for mitigating
entanglement dissipation.  We first consider a simple scheme based on
open-loop coherent control and then a measurement-based feedback scheme.

\subsection{Coherent quantum control}

For pure dephasing and a hamiltonian of the form (\ref{ham}) the
$(2,3)$-subspace is invariant, i.e. if a state starts in this subspace
it remains there.  Thus if the system is initially prepared in the Bell
state Eq.~(\ref{bell}) then we may consider the reduced dynamics on
the $(2,3)$-subspace.  The density operator $\rho$ restricted to the
$(2,3)$-subspace is essentially a qubit, i.e., a $2 \times 2$ positive
matrix $\tilde{\rho}$.  Introducing the Pauli matrices
\begin{equation}\label{Pauli}
  X=\left[\begin{array}{cc}0&1\\1&0  \end{array}\right],\; \;
  Y=\left[\begin{array}{cc}0&-i\\i&0 \end{array}\right],\; \;
  Z=\left[\begin{array}{cc}1&0\\0&-1 \end{array}\right]
\end{equation}
then the Lindblad equation Eq(\ref{lin}), with a single $V$ term $=Z$,
takes the simple form
\begin{equation}
 \dot{\tilde{\rho}} = -i [y X, \tilde{\rho}] + \DD(Z)\tilde{\rho}
\end{equation}
where $\DD(Z)\tilde{\rho}=Z \tilde{\rho} Z-\tilde{\rho}$.

It is convenient to use the Bloch representation, for which the 3-vector
$\vec{s}$ associated with the density matrix $\rho$ is given by
\begin{equation}
 \label{Bloch}
  \vec{s}= ({\rm trace}(X \tilde{\rho}),{\rm trace}(Y \tilde{\rho}),{\rm trace}(Z \tilde{\rho}))
\end{equation}
For hamiltonian dynamics, $\norm{\vec{s}}$ is constant and the motion is
governed by the orthogonal group $O(3)$ acting on the Bloch Sphere.  In
the case of dissipation, the motion takes place in its interior, the
Bloch Ball, and is locally governed by the affine Lie algebra
$gl(3)\oplus R^3$; this is globally a semi-group due to boundary
conditions~\cite{paris}.  Therefore in general the Bloch equation is
\begin{equation}
 \label{Bloch eq}
   \dot{\vec{s}}(t) = A\vec{s}(t)+\vec{c}.
\end{equation}
In the simple case considered here, we have
\begin{equation}
  A = \begin{bmatrix}
      -2 &  0 &   0\\
       0 & -2 & -2y\\
       0 & 2y &   0
    \end{bmatrix}, \quad
\vec{c} = \begin{bmatrix} 0 \\ 0 \\ 0 \end{bmatrix}.
\end{equation}
For $y \neq 0$ the Bloch matrix $A$ is invertible and any initial state
eventually goes to the unique steady state $\vec{s}_{\ss}=-A^{-1}
\vec{c}=[0,0,0]^T$, which corresponds to the completely mixed state on
the $(2,3)$-subspace, which has zero concurrence.  Thus regardless of
the value of the control $y$ the system will eventually go to the
completely mixed state on the $(2,3)$-subspace and all entanglement will
be lost.  If $y=0$ (no control) then all the Bloch vectors along the
$z$-axis in the $(2,3)$-subspace are steady states, but if we start with
a Bell state we still go to the completely mixed state on the subspace
for $t\to\infty$.  {\em So in this case there is no way coherent control
can prevent the decay of the entanglement.}  This is understandable as the
hamiltonian term determines the anti-symmetric part of $A$ while the
decoherence term defines the symmetric part, so the control cannot
change the contraction of the Bloch vector introduced by the dephasing
term.  If we want to stabilize an entangled state we need to employ a
more sophisticated method.  One approach is to utilize a measurement and
feedback scheme.

\subsection{Measurement/feedback control scheme}

One approach that has shown recent promise in particular for stabilizing
quantum states~\cite{schirmer} is reservoir engineering, in particular
using direct feedback.

Suppose we have a system whose evolution is governed by a Lindblad
master equation~(\ref{eq:dLE}).  If we add a fixed continuous weak
measurement of an observable $M$ and apply a fixed feedback hamiltonian
$F$ conditioned \emph{directly} on the measurement current, then
according to the general theory developed by Wiseman and Milburn~\cite{wiseman,wm}
the master equation is modified as follows:
\begin{equation}
  \label{fmeq}
  \dot{\rho} = -i [H,\rho] + \DD[M-i F]\rho + \L_D(\rho)
\end{equation}
where $H$ is the original system hamiltonian plus the open-loop control
hamiltonian $H_0+H_c$ plus a feedback correction term $(M^\dag F + F
M)/2$ (see the diagrammatic scheme of Figure 3).

\begin{figure}
\includegraphics{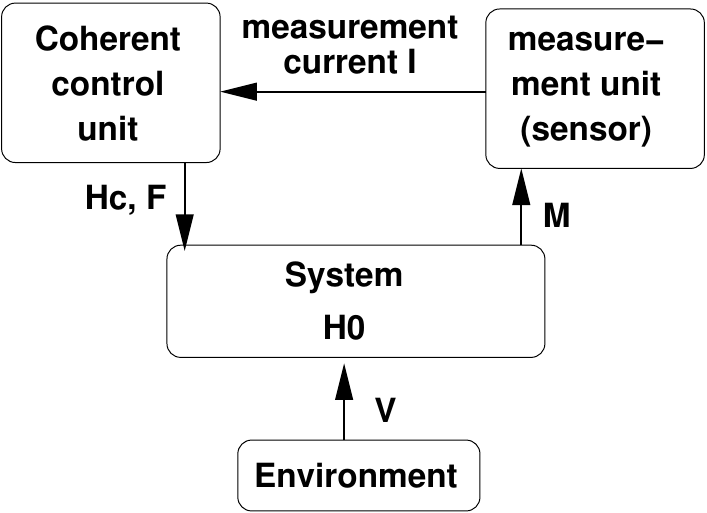}
\caption{Basic direct-feedback control setup.}
\end{figure}

Suppose we have a system with hamiltonian $H_0$ as in (\ref{ham})
subject to environmental dephasing described by a diagonal Lindblad
operator $V$.  If we now add a weak continuous measurement of the
observable $\sqrt{m} Z\otimes I$ and apply a feedback hamiltonian of the
form $\sqrt{f} X\otimes X$ then the evolution of the system according
to (\ref{fmeq}) is governed by
\begin{equation}
  \dot{\rho} = -i [H,\rho] + + \DD[V](\rho) + \DD[M-i F]\rho
\end{equation}
As observed before, the $(2,3)$ subspace is invariant under $H$ and $V$.
It is also invariant under $M-iF$, i.e. any state starting in this
subspace will remain in it.  Thus, if we are only interested in initial
states in this subspace we can again restrict our attention to the
dynamics on this subspace.  The subspace operators are
\begin{equation}
  H_0^{(2,3)} = \mu Z + y X, \quad
  M^{(2,3)} = \sqrt{m}Z, \quad
  F^{(2,3)} = \sqrt{f}X, \quad
  V^{(2,3)} = \sqrt{\Gamma} Z
\end{equation}
and noting that $Z^\dag X + X Z=0$ shows that the master equation
for the density operator $\tilde{\rho}$ restricted to the $(2,3)$
subspace takes the form
\begin{equation}
 \dot{\tilde{\rho}} = -i [y X + \mu Z, \tilde{\rho}] +
  \DD[\sqrt{m} Z-i \sqrt{f} X] \tilde{\rho}
                + \Gamma \DD[Z] \tilde{\rho}.
\end{equation}
Here $\mu$ is the effective energy level splitting and $\Gamma$ the
effective environmental decoherence rate, and $y$, $\sqrt{m}$ and
$\sqrt{f}$ are the effective strengths of the open-loop control,
measurement and feedback, respectively.  The corresponding Bloch
operator is
\begin{equation}
 A=-2\, \begin{pmatrix}
         m+\Gamma&\mu&0 \\
	 -\mu&f+m+\Gamma&y\\
	 0&-y&f
\end{pmatrix}
\end{equation}
and $\vec{c}=[0,4\sqrt{mf},0]^T$.

For calculational simplicity we choose the open-loop control term to be
$0$, i.e., $y=0$.  In this case it is easy to compute the eigenvalues of
$A$:
\begin{equation}
 -2f, -f-2\Gamma-2m \pm \sqrt{f^2-4\mu^2}.
\end{equation}
Thus $A$ is non-degenerate if $f\neq 0$ and the system therefore has
a unique steady state $\vec{s}_{\ss}=-A^{-1} \vec{c}$ on the subspace,
which in density operator form is
\begin{equation}
  \tilde{\rho}_{\ss} =
   \begin{bmatrix}
  \frac{1}{2} & \frac{\sqrt{f m}(\mu+i(\Gamma+m))}{\mu^2+(\Gamma+m)(\Gamma+m+f)} \\
  \frac{\sqrt{f m}(\mu-i(\Gamma+m))}{\mu^2+(\Gamma+m)(\Gamma+m+f)} & \frac{1}{2}
   \end{bmatrix}.
\end{equation}
The purity of the steady state is given by
\begin{equation}
  P_{\ss} = \frac{1}{2} + 2
  \frac{f m (\mu^2+(\Gamma+m)^2)}{(\mu^2+(\Gamma+m)(\Gamma+m+f))^2}
\end{equation}
and the concurrence is
\begin{equation}
  C_{\ss} = 2\frac{\sqrt{mf}(\mu^2+(\Gamma+m)^2)^{1/2}}{\mu^2+(\Gamma+m)(\Gamma+m+f)} \ge 0.
\end{equation}
Note that $C_{\ss} = \sqrt{2P_{\ss}-1}$, i.e. there is a one-to-one
correspondence between the concurrence and purity in this case.  If we
further choose $\mu=0$ (adjust energy level splitting to be $0$) then
the expressions simplify:
\begin{equation}
 \tilde{\rho}_{\ss} = \begin{pmatrix}
	       \frac{1}{2} & i \frac{\sqrt{f m}}{\Gamma+f+m}\\
	       -i\frac{\sqrt{f m}}{\Gamma+f+m} & \frac{1}{2}
	      \end{pmatrix}
\end{equation}
and the purity and concurrence are given by
\begin{align}
  P_{\ss} &= \frac{1}{2} + \frac{2f m}{(\Gamma+f+m)^2} \\
  C_{\ss} &= 2 \frac{\sqrt{mf}}{\Gamma+m+f}.
\end{align}

\begin{figure}[t]
\includegraphics[width=1.0\textwidth]{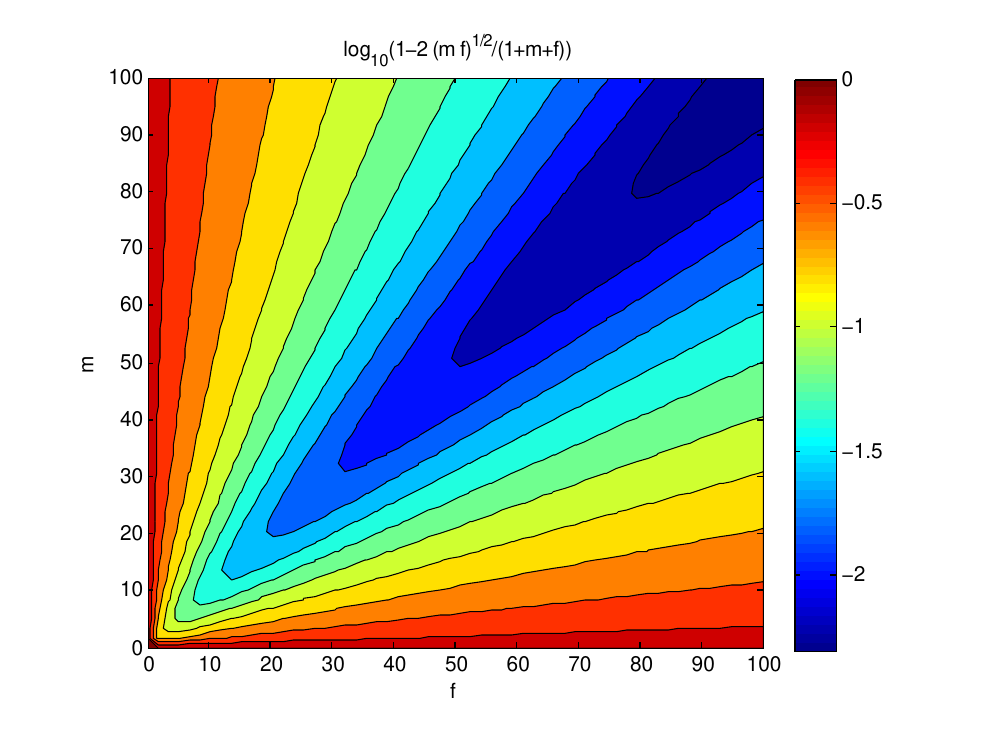}
\caption{Steady-state concurrence $C_{\ss}$, or precisely
$\log_{10}(1-C_{\ss})$, as a function of the measurement and feedback
strength $m$ and $f$, respectively, assuming $\Gamma=1$.}
\label{fig:conc}
\end{figure}

For a given fixed decoherence rate $\Gamma>0$, we can choose $m$ and
$f$ in units of $\Gamma$ and set $\Gamma=1$.  In this case the
steady-state concurrence becomes
\begin{equation}
  C_{\ss} = 2 \frac{\sqrt{mf}}{1+m+f}.
\end{equation}
Figure~\ref{fig:conc} shows the logarithm of $1-C_{\ss}$ as a function of
the measurement and feedback strengths.  We see that the optimum choice
to maximize the concurrence is $m=f$ with $m$ and $f$ as large as
practically feasible.  The plot given in Figure 4 shows that if the measurement and
feedback strengths are about 100 times the environmental decoherence rate
$\Gamma$ then the steady-state concurrence is greater than $1-10^{-2}$.

Although this scheme may be difficult to realize in practice because we
require a non-local feedback hamiltonian of the form $X\otimes X$, with some
experimental ingenuity such a procedure may well be implemented.

.

\end{document}